\documentclass[sigconf]{acmart}

\usepackage{ragged2e} 
\usepackage{booktabs,makecell, multirow, tabularx}
\usepackage{bm}
\usepackage{makecell}
\usepackage{graphicx}
\usepackage{balance}

\AtBeginDocument{%
  }

\copyrightyear{2024}
\acmYear{2024}
\setcopyright{acmlicensed}
\acmConference[MM '24]{Proceedings of the 32nd ACM International Conference on Multimedia}{October 28-November 1, 2024}{Melbourne, VIC, Australia}
\acmBooktitle{Proceedings of the 32nd ACM International Conference on Multimedia (MM '24), October 28-November 1, 2024, Melbourne, VIC, Australia} 
\acmDOI{10.1145/3664647.3688988}
\acmISBN{979-8-4007-0686-8/24/10}

\begin{document}

\title[Dialogue-Aware Transformer for Engagement Estimation]{DAT: Dialogue-Aware Transformer with Modality-Group Fusion for Human Engagement Estimation}


\author{Jia Li}
\authornote{Equally contribution.}
\affiliation{%
  \institution{School of Computer Science and Information Engineering, Hefei University of Technology}
  \city{Hefei}
  \country{China}}
\email{jiali@hfut.edu.cn}

\author{Yangchen Yu}
\authornotemark[1]
\affiliation{%
  \institution{School of Computer Science and Information Engineering, Hefei University of Technology}
  \city{Hefei}
  \country{China}}
\email{2019212292@mail.hfut.edu.cn}

\author{Yin Chen}
\affiliation{%
  \institution{School of Computer Science and Information Engineering, Hefei University of Technology}
  \city{Hefei}
  \country{China}}
\email{chenyin@mail.hfut.edu.cn}

\author{Yu Zhang}
\affiliation{%
  \institution{School of Computer Science and Information Engineering, Hefei University of Technology}
  \city{Hefei}
  \country{China}}
\email{yuyueback@gmail.com}

\author{Peng Jia}
\affiliation{%
  \institution{School of Computer Science and Information Engineering, Hefei University of Technology}
  \city{Hefei}
  \country{China}}
\email{2020214631@mail.hfut.edu.cn}

\author{Yunbo Xu}
\affiliation{%
  \institution{School of Computer Science and Information Engineering, Hefei University of Technology}
  \city{Hefei}
  \country{China}}
\email{xuyunbocn@mail.hfut.edu.cn}

\author{Ziqiang Li}
\affiliation{%
  \institution{School of Computer Science and Information Engineering, Hefei University of Technology}
  \city{Hefei}
  \country{China}}
\email{ziqlee1029@gmail.com}

\author{Meng Wang}
\affiliation{%
  \institution{School of Computer Science and Information Engineering, Hefei University of Technology}
  \city{Hefei}
  \country{China}}
\email{eric.mengwang@gmail.com}

\author{Richang Hong}
\authornote{Corresponding author.}
\affiliation{%
  \institution{School of Computer Science and Information Engineering, Hefei University of Technology}
  \city{Hefei}
  \country{China}}
\email{hongrc.hfut@gmail.com}
\renewcommand{\shortauthors}{Jia Li et al.}

\begin{abstract}
Engagement estimation plays a crucial role in understanding human social behaviors, attracting increasing research interests in fields such as affective computing and human-computer interaction. In this paper, we propose a Dialogue-Aware Transformer framework (DAT) with Modality-Group Fusion (MGF), which relies solely on audio-visual input and is language-independent, for estimating human engagement in conversations. Specifically, our method employs a modality-group fusion strategy that independently fuses audio and visual features within each modality for each person before inferring the entire audio-visual content. This strategy significantly enhances the model's performance and robustness. Additionally, to better estimate the target participant's engagement levels, the introduced Dialogue-Aware Transformer considers both the participant's behavior and cues from their conversational partners. Our method was rigorously tested in the Multi-Domain Engagement Estimation Challenge held by MultiMediate'24, demonstrating notable improvements in engagement-level regression precision over the baseline model. Notably, our approach achieves a CCC score of 0.76 on the NoXi Base test set and an average CCC of 0.64 across the NoXi Base, NoXi-Add, and MPIIGI test sets. The source code will be available at \url{https://github.com/MSA-LMC/DAT}.

\end{abstract}

 



\maketitle

\section{Introduction}
In communication, engagement is a crucial indicator of participants' involvement and interaction levels within a conversation. It not only reflects the importance participants attribute to the interaction but also profoundly influences the quality and efficacy of the dialogue. The rapid advancements in deep learning technologies have led to the emergence of automating engagement estimation as a critical challenge and opportunity in affective computing and group behavior analysis \cite{zheng2021estimation, kaur2018prediction}.

To further research in this field, the MultiMediate 2024 competition has organized the Multi-Domain Engagement Estimation track \cite{muller2024multimediate}. Competitors are tasked with continuously predicting the engagement level of each individual in a conversation, frame by frame, with scores ranging from 0 to 1. Competitors are encouraged to explore multimodal data and reciprocal behaviors, and to evaluate the accuracy of their predictions using the Concordance Correlation Coefficient (CCC) \cite{lawrence1989concordance}.

\begin{figure}[h]
  \centering
  \includegraphics[width=\linewidth]{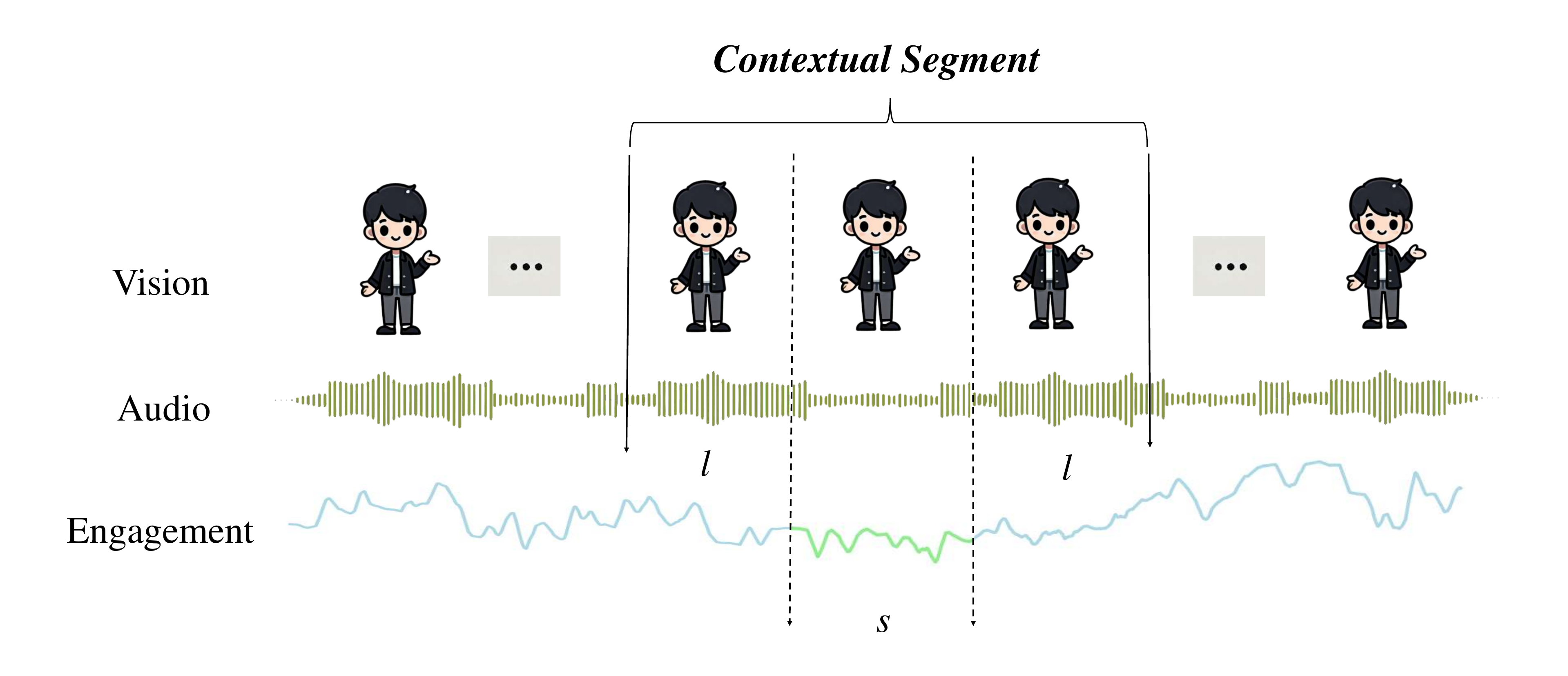}
  \caption{The task of engagement estimation \cite{muller2024multimediate}  aims to predict the continuous values of a target participant's engagement level frame-by-frame in a conversation. Contextual segmentation is performed on the entire recording sequence to include more contextual information around a specific time period for prediction.}
  \label{fig:sw2}
\end{figure}

Existing methods for engagement estimation typically rely on extracting features from video sequences or segments, followed by processing them with Recurrent Neural Networks (RNN) \cite{elman1990finding}, Long Short-Term Memory networks (LSTM) \cite{hochreiter1997long}, or Transformer \cite{vaswani2017attention} models. These approaches focus primarily on utilizing cues from the target participants, overlooking the conversational partner's behavior, which also plays a crucial role in evaluating the target participant's engagement. Moreover, in multimodal engagement estimation tasks, current methods often simply concatenate features extracted from video sequences or segments for engagement prediction, without deeply exploring the interplay between various features and modalities.

To address these limitations and challenges, we propose a flexible and effective framework, named DAT, which is designed to fully utilize the audio and visual information of relevant participants in the conversation to accurately estimate the engagement level of the target participant. Specifically, DAT is composed of several Transformer Layers, including the Modality-Group Fusion (MGF) module and the Dialogue-Aware Transformer Encoder (DAE). The MGF module aims to encode video and audio features using a modality-group fusion strategy to obtain deep and language-independent modality-specific representations. This unique processing of modalities aims to minimize redundancy and overlap within the datasets. Additionally, the DAE module, a straightforward Transformer with Cross-Attention, incorporates auxiliary information from other conversational partners to enrich the target participant's data. The enrichment of data sources enhances the precision in capturing the target participant's behavior throughout the conversation, thereby enhancing the accuracy of the engagement estimation. 

Our method was evaluated on the NoXi Base \cite{cafaro2017noxi, muller2024multimediate} , NoXi-Add \cite{muller2024multimediate}, and MPIIGI \cite{muller2018detecting} test sets in the \textit{Multi-Domain Engagement Estimation Challenge of the MultiMediate 2024}. The experimental results indicate that the proposed method achieved superior performance and surpassed baseline models ~\cite{muller2024multimediate} across all datasets in terms of CCC, confirming that the inclusion of conversational partners and the modality-group fusion strategy significantly enhances our model performance and robustness.

The remainder of this paper is organized as follows. Related works are introduced in Section \ref{related_work}. Section \ref{ourmethod} describes the details of our DAT model and the whole workflow. Section \ref{experiments} presents the implementation details, ablation studies and comparisons with other methods. We conclude our work in Section \ref{conclusion}.

\section{Related Works}
\label{related_work}
\textbf{Engagement Estimation}. Estimating engagement \cite{10.1145/3503161.3548363} levels  in human conversation remains a pivotal topic in human-computer interaction. This task entails predicting participants' engagement frame-by-frame, utilizing multimodal data from interactions involving two or more individuals. Several studies \cite{lala2017detectionsocialsignalsrecognizing, guhan2023developingeffectiveautomatedpatient,bednarik2012gaze,sanghvi2011automatic,yu2023sliding} have explored engagement estimation. Additionally, studies within the education sector \cite{goldberg2021attentive,karimah2021automatic,saleh2022video} have also addressed this topic. Recent efforts have increasingly directed attention towards estimating engagement within conversational settings. Yang et al. \cite{10.1145/3581783.3612873}, achieved commendable results by classifying and concatenating audio-video features using a straightforward network architecture. Similarly, Yu et al. \cite{10.1145/3581783.3612852} employed a sliding window approach to capture both local and global cues in video data, utilizing Transformer's capabilities for sequence modeling to depict participant engagement accurately. However, these methodologies insufficiently explore the multimodal data inherent in conversations, thus missing out on capturing the nuanced representations of interaction data. Our research, in contrast, emphasizes dialogue scenarios and has shown good performance across multiple engagement estimation datasets.

\noindent \textbf{Multimodal Feature Fusion}. To enhance engagement estimation, it is crucial to employ multimodal feature fusion techniques to bridge the semantic gap between different modalities. Several approaches have been proposed to leverage different modalities. Commonly, self-attention \cite{vaswani2017attention} and cross-attention \cite{tan2019lxmertlearningcrossmodalityencoder} mechanisms are employed. Beyond traditional self-attention and cross-attention mechanisms, more sophisticated techniques exist. Neural Multimodal Embeddings \cite{caglayan2016multimodalattentionneuralmachine} learn a joint embedding space mapping different modalities to proximate, semantically similar points, facilitating tasks such as cross-modal retrieval, generation, and alignment. Tensor Fusion \cite{zadeh2017tensorfusionnetworkmultimodal} addresses both inter- and intra-modality aspects by considering interactions within and between modalities.
Inspired by prior work \cite{10.1145/3581783.3612873} , our strategy partitions features into video and audio groups before their subsequent fusion, achieving better results.

\begin{figure*}[!ht]
  \centering
  \includegraphics[width=0.96\linewidth]{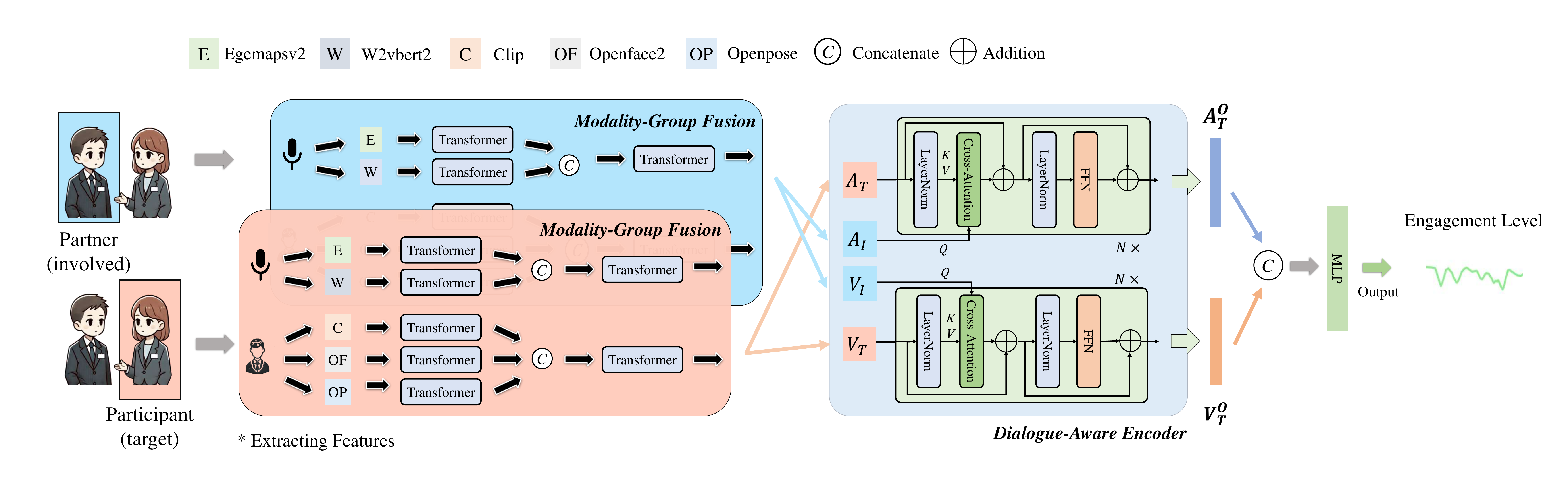}
  \caption{Overall architecture of the proposed method.
  \textnormal{Our DAT consists of two main modules: Modality-Group Fusion and Dialogue-Aware Encoder. Firstly, the Modality-Group Fusion module processes audio and visual features for both the participant and partner. Each feature is processed through a Transformer before being fused together. Subsequently, the Dialogue-Aware Encoder utilizes cross-attention to combine and encode information from both participants, focusing on contextual interactions to enhance engagement prediction. Finally, an MLP predicts continuous engagement levels frame-by-frame by utilizing the encoded features.}} 
  \label{fig:model_noxibase}
\end{figure*}

\noindent \textbf{Emotion Analysis}. The problem of human engagement estimation is highly correlated with the popular research field of human emotion analysis, such as facial expression recognition in images or videos, audio-visual emotion recognition \cite{zadeh2017tensor, li2022emotion,ning2024representation}, and multimodal sentiment analysis \cite{chen2023static, tellamekala2023cold,chumachenko2024mma}. Emotion expressed by a person's facial expression, speech tone and spoken language is closely related to their level of engagement in a dialogue. Considering that large-scale unimodal or multimodal in-the-wild emotion datasets have been established, we believe that emotion-related audio, visual, and textual features (or representations) can benefit existing methods for engagement estimation if only the original multimodal data of conversations is available. However, in the MultiMediate'24 Challenge, only pre-extracted specific features are provided, and the original data is not available.

\section{Methodology}
\label{ourmethod}

\subsection{Problem Definition}


Given a long video, the engagement estimation task involves predicting the continuous level of conversational engagement for each participant, ranging from 0 (minimal engagement) to 1 (maximum engagement). Since each frame in the video is individually labeled, this task is essentially a continuous regression problem. Given the inputs of target participant $\bm{X}$ and the label of the target participant $Y$, this task is to learn a mapping function $f_\theta(\bm{X}) \rightarrow \bm{Y}$, where $f_{\theta}$ defines the model and $\theta$ represents the learnable parameters.

\subsection{Contextual Segment}
For the engagement estimation problem, the ideal situation is that the engagement of the participants at the current moment can be predicted by combining the features of the current moment frame and the frames within a period of time around it, as illustraed in Figure \ref{fig:sw2}. This problem can be formalized as follows:
\begin{equation}
  y_{t} = f_{\theta}(x_{t - l},x_{t - l + 1},x_{t - l + 2}, \ldots ,x_{t} ,\ldots ,x_{t + l - 2},x_{t + l -1},x_{t + l}),
\end{equation}
where $[ x_{t - l}, x_{t - l + 1}, x_{t - l + 2}, \ldots, x_{t + l}]$ represents the multimodal features from $t - l$ to $t + l$ time, $y_t$ represents the estimated engagement of the participant at time $t$, $l$ is the length of the surrounding frames considered.



Contextual segment-based prediction can significantly enhance performance beyond what is achievable through the sole use of multimodal features at any given moment. Engagement levels of participants are in constant flux throughout a conversation, where the contextual backdrop often holds essential insights crucial for accurately estimating engagement. To leverage this, we introduce context segmentation for the input feature sequence, i.e., we divide the 
video into several sub-sequences tailored for subsequent prediction tasks.  our approach enables simultaneous engagement prediction across multiple frames, employing a step size $s$ to generate overlapping sub-sequences. These overlapping sub-sequences are processed using a Sequence-to-Sequence (Seq2Seq) Model, which is delineated as follows:
\begin{equation}
  y_{0},\cdots , y_{s} = f_{\theta}(x_{-l},\cdots ,x_{-1},x_{0},\cdots,x_{s},x_{s + 1},\cdots,x_{s + l}).
\end{equation}

Here, $s + 2l$ denotes the sequence length of each prediction input, where $s$ is the predicted sequence length (stride), and $x_{i}$ represents the $i$ multimodal feature in the sequence. A larger $l$ incorporates more surrounding frames. When $l$ is sufficiently large, the entire video can be approximately modeled.

We observe that certain situations within a conversation, such as unexpected phone calls, can cause significant fluctuations in participants' engagement levels. Additionally, longer sequences tend to contain more extraneous information, which can negatively impact model performance. Thus, selecting an appropriate sequence length is crucial for effectively capturing relevant features and enhancing prediction accuracy.
 
\subsection{DAT Model}


As illustrated in Figure \ref{fig:model_noxibase}, a Dialogue-Aware Transformer (DAT) framework has been developed specifically for engagement estimation. The DAT framework comprises Modality-Group Fusion (MGF) modules and a Dialogue-Aware Encoder (DAE). Each MGF module independently processes the video and audio features from both the target participant and their conversational partner, and the DAE combines these features, facilitating their effective integration. The resulting fused feature is then feed to MLP to predict the final engagement level of the target participant within the conversation.

\subsubsection{\textbf{Modality-Grouping Fusion}}
As depicted in Figure \ref{fig:model_noxibase}, five features ( $\bm{X}_E$, $\bm{X}_W$, $\bm{X}_C$, $\bm{X}_{OF}$ and  $\bm{X}_{OP}$ ) are extracted from the  conventional participants using several specialized tools: OpenSmile ~\cite{eyben2010opensmile}, W2v-BERT 2.0 ~\cite{chung2021w2v}, CLIP ~\cite{radford2021learning}, OpenFace ~\cite{baltruvsaitis2016openface} and OpenPose ~\cite{cao2017realtime}, respectively. Initially, these features are  segmented into several contextual  segments according to their timestamps. For each segment feature $\bm{x}_a$, $a\in \{E, W,C, OF, OP\}$, it is first projected  to a unified dimension $d$ then processed though a Transformer to derive deeper representations. The mathematical formulation of this process is as follows:
\begin{equation}
    \bm{x}_a'=Proj(\bm{x}_a), a\in \{E, W, C, OF, OP\},
\end{equation}
\begin{equation}
    \bm{x}_a''=Transformer(\bm{x}_a'),
\end{equation}
where $Proj$  denotes the linear layer and the ${Transformer}$ represents a standard transformer layer.

Subsequently, these deep features are grouped into audio features $\bm{x}''_E$, $\bm{x}''_W \in \mathbb{R}^{(s+2l) \times d}$ and video feature $\bm{x}''_C$, $\bm{x}''_{OF}$, $\bm{x}''_{OP} \in \mathbb{R}^{(s+2l) \times d}$, which are then concatenated in $\bm{x}_A \in \mathbb{R}^{(s+2l) \times 2d}$, $\bm{x}_V \in \mathbb{R}^{(s+2l) \times 3d}$. This can be described as:
\begin{equation}
    \bm{x}_A = Concat(\bm{x}_E'', \bm{x}_W''),
\end{equation}
\begin{equation}
    \bm{x}_V = Concat(\bm{x}_C'', \bm{x}_{OF}'', \bm{x}_{OP}''), 
\end{equation} 
where $Concat$ denotes concatenation operation. Following this, the audio features and video features are further processed using a Transformer layer to derive deeper representations of audio and video modalities:
\begin{equation}
    \bm{x}_A' = Transformer(\bm{x}_A),
\end{equation}
\begin{equation}
    \bm{x}_V' = Transformer(\bm{x}_V),
\end{equation}
where $\bm{x}_A'\in \mathbb{R}^{(s+2l) \times 2d}$ and $\bm{x}_V'\in \mathbb{R}^{(s+2l) \times 3d}$ represent the processed audio and video modality features for the participant, respectively.

\subsubsection{\textbf{Dialogue-Aware Transformer Encoder}}
To enhance the accuracy of estimating the participation level of individuals in a conversation, relying solely on the data from the target participant is insufficient and biased. Consequently, this paper advocates for the inclusion of the conversational partner's information as supplementary data. We designed a dialogue-aware transformer encoder to fuse the audio and video features from the target participant and involved partners. As illustrated in Figure \ref{fig:model_noxibase}, the modality features $\bm{A}_T$, $\bm{V}_T$ for target participant and $\bm{A}_I$ , $\bm{V}_I$ for involved partner are obtained from the Modality-Group Fusion Module, respectively. Upon extracting these features, the partner's features are used as the query, and the target participant's features as both the key and value, to perform cross-attention fusion on audio and video features independently. This is achieved through $N$ layers of Transformer to improve the data integration process. The dialogue-aware transformer encoder layer's operation is as follows:
\begin{equation}
    \bm{A}'_T = Norm(\bm{A}_T),
\end{equation}
\begin{equation}
    \bm{A}''_T = Cross\text{-}Attention(\bm{A}_I,\bm{A}'_T, \bm{A}'_T) + \bm{A}_T,
\end{equation}
\begin{equation}
        Cross\text{-}Attention(\bm{Q},\bm{K},\bm{V}) = softmax \left( \frac{\bm{QK}^\top}{\sqrt{d_k}} \right) \bm{V},
\end{equation}
where $Norm$, $FFN$, $softmax$ denote the LayerNorm and FeedForward and softmax function, respectively. The final audio feature output is obtained as follows:
\begin{equation}
    \bm{A}^O_T =  \bm{A}''_T + FFN(Norm(\bm{A}''_T)).
\end{equation}
Similarly, the final video feature $\bm{V}^O_T$ is derived using dialogue-aware transformers.

This methodology not only helps to filter out irrelevant information but also allows the model to consider a broader and more comprehensive dataset, enabling a more accurate evaluation of the target participant’s engagement level.

For all the encoders of the aforementioned models, we utilize fully connected layers to predict the final engagement level of the target participant. The final prediction $\bm{y}^{pre} $ is calculated as follows:
\begin{equation}
        \bm{x}^O_T = Concat(\bm{A}^O_T, \bm{V}^O_T),
\end{equation}

\begin{equation}
    \bm{y}^{pre}  = MLP(\bm{x}^O_T),
\end{equation}
where $\bm{x}^O_T \in \mathbb{R}^{(s+2l) \times 5d}$ and $MLP$ denotes fully connected layers.

Finally, we supervise the model using MSE loss for the NoXi datasets and CCC loss for the MPIIGroupInteraction (MPIIGI) dataset. Different loss functions are employed due to the distinct characteristics of their labels. The labels in the NoXi dataset are continuous and fluctuate between adjacent frames, making MSE loss suitable for capturing these variations. In contrast, the MPIIGI dataset labels are discrete, categorized into 25 classes ranging from 0 to 1, resulting in many consecutive frames having identical labels. Therefore, CCC loss is more appropriate for MPIIGI as it effectively handles the stability and consistency of discrete labels across multiple frames.

\section{Experiments}
\label{experiments}
To demonstrate the efficacy of our proposed DAT framework in engagement estimation tasks, we conduct extensive evaluations on the dataset provided by the organizers. Moreover, ablation studies are conducted on Modality-Group fusion module and Dialogue-Aware Transformer encoder. This section is dedicated to detailing the datasets employed during both training and evaluation stages, outlining the experimental setup, presenting an ablation analysis, and highlighting the most favorable experimental results.

\subsection{Datasets}
This challenge ~\cite{muller2024multimediate} includes three datasets, and the average CCC score ~\cite{lawrence1989concordance} across these datasets is used as the final competition metric.

\textbf{NoXi Base ~\cite{cafaro2017noxi} dataset} includes 38 sets of training files, 10 sets of validation files, and 16 sets of test files. It primarily consists of one-on-one video interactions between experts and novices in English, French, and German, incorporating both audio and visual components. The sampling rate is 25 fps, although some features are sampled at 40 fps. Due to privacy concerns, only pre-extracted features are provided. These features are continuously recorded, with each frame labeled with an engagement value ranging from 0 to 1.

\textbf{NoXi Additional Language dataset \textnormal{(NoXi-Add)}} comprises only 12 sets of test files. Unlike the basic version of NoXi, the linguistic forms in its features are entirely different, including Arabic, Italian, Indonesian, and Spanish. However, other feature formats remain consistent with the basic version.

\textbf{MPIIGroupInteraction ~\cite{mueller21_mm} dataset} consists of 6 validation files and 6 test files. MPIIGI dataset records the engagement levels of four individuals interacting in real-time using German. Four cameras capture the individuals from different positions. A significant difference from the NoXi dataset is that participants in MPIIGI dataset are mostly recorded in a seated position, whereas those in NoXi dataset are recorded standing. The sampling rates of various features in MPIIGI dataset remain consistent with those in the NoXi dataset.

\subsection{Experimental Setup} 


We utilized all official-provided features from the dataset, totaling 2477 dimensions, including: 88-dimensional audio features from OpenSmile ~\cite{eyben2010opensmile}, 1024-dimensional word features from W2v-BERT 2.0 ~\cite{chung2021w2v}, 512-dimensional visual features from CLIP ~\cite{radford2021learning}, 714-dimensional facial features from OpenFace ~\cite{baltruvsaitis2016openface}, and 139-dimensional pose features from OpenPose ~\cite{cao2017realtime}. These features are grouped into audio (first two) and visual (latter three) modalities. 

In the contextual segment method, the total sequence length is 96, with a core sequence length of 32 and auxiliary sequence lengths of 32 on both sides. The step size $s$ for segment division is also 32. Additionally, all features in Modality-Group Fusion are uniformly mapped to a dimension $d$ of 512. As shown in Figure~\ref{fig:model_noxibase}, the number of layers $N$ for both modalities in the Dialogue-Aware Encoder is set to 1. To reduce overfitting and improve model generalization, the dropout rate for all network layers is set to 0.2.

To optimize the model, we utilize the Adam optimizer with a learning rate of 5e-5 and a batch size of 128. The training process incorporates different loss functions tailored to each dataset: Mean Squared Error (MSE) for the NoXi dataset and Concordance Correlation Coefficient (CCC) for the MPIIGI dataset. Additionally, an Exponential Moving Average (EMA) strategy is applied throughout the 50 training epochs. For the NoXi Base and NoXi-Add datasets, the NoXi train and validation sets are used as the training data, while the MPIIGI validation set serves as the training data for the MPIIGI dataset.

\subsection{Ablation Study}


This section investigates the impact of our proposed key modules, Modality-Group Fusion and Dialogue-Aware Encoder, on model performance through ablation experiments on the NoXi Base dataset. Notably, our baseline model does not benefit from these two modules. It consists of six Transformer encoders ~\cite{vaswani2017attention}, where five features are sequentially enhanced through five encoders, concatenated, and integrated through one Transformer. Predictions are made through the MLP layers. This corresponds to the first row in Table~\ref{tab:tab1}, representing our model's baseline.

\noindent \textbf{Ablation on Modality-Grouping Fusion Module.}
We have developed a feature fusion strategy and introduced the Modality-Group Fusion (MGF) module. Our hypothesis suggests that by grouping and fusing features based on their respective modalities, it is possible to minimize information redundancy within each modality, thereby optimizing inter-modality integration. Consequently, the MGF module was specifically designed to group and fuse audio and video features, thereby enhancing the overall feature fusion process. The results as presented in Table~\ref{tab:tab1} clearly support our hypothesis. The integration of the MGF module has resulted in a performance enhancement by 0.014 on the validation set. This improvement further substantiates the effectiveness of the MGF module in boosting the model's learning capabilities and enhancing its overall performance.

\begin{table}
  \caption{Ablation study on Modality-Grouping Fusion Module (MGF) and Dialogue-Aware Encoder (DAE) on the NoXi Base dataset.}  
  \label{tab:tab1}
  \scalebox{0.88}{
  \begin{tabular}{ccc}
    \toprule
    MGF&DAE&Val CCC\\
    \midrule
    $\times$& $\times$ & 0.818\\
    $\times$& \checkmark & 0.826\\
    \checkmark& $\times$ & 0.832\\
    \checkmark& \checkmark & 0.846\\
  \bottomrule
\end{tabular}}
\end{table}

\begin{table}
\centering
\caption{Ablation study controlling the parameter count with and without the Dialogue-Aware Encoder (DAE) on the NoXi Base dataset. Encoder Depth refers to setting all Transformer layers in the model to the same value.}
\scalebox{0.88}{
\begin{tabular}{cccc}
\toprule
DAE & Encoder Depth & Tunable Params (M) & Val CCC \\
\midrule
$\times$ & 2 & 112 & 0.821 \\
\checkmark & 1 & 113 & 0.846 \\
\bottomrule
\end{tabular}}
\label{your-label}
\end{table}

\begin{figure}
    \centering
    \includegraphics[width=0.7\linewidth]{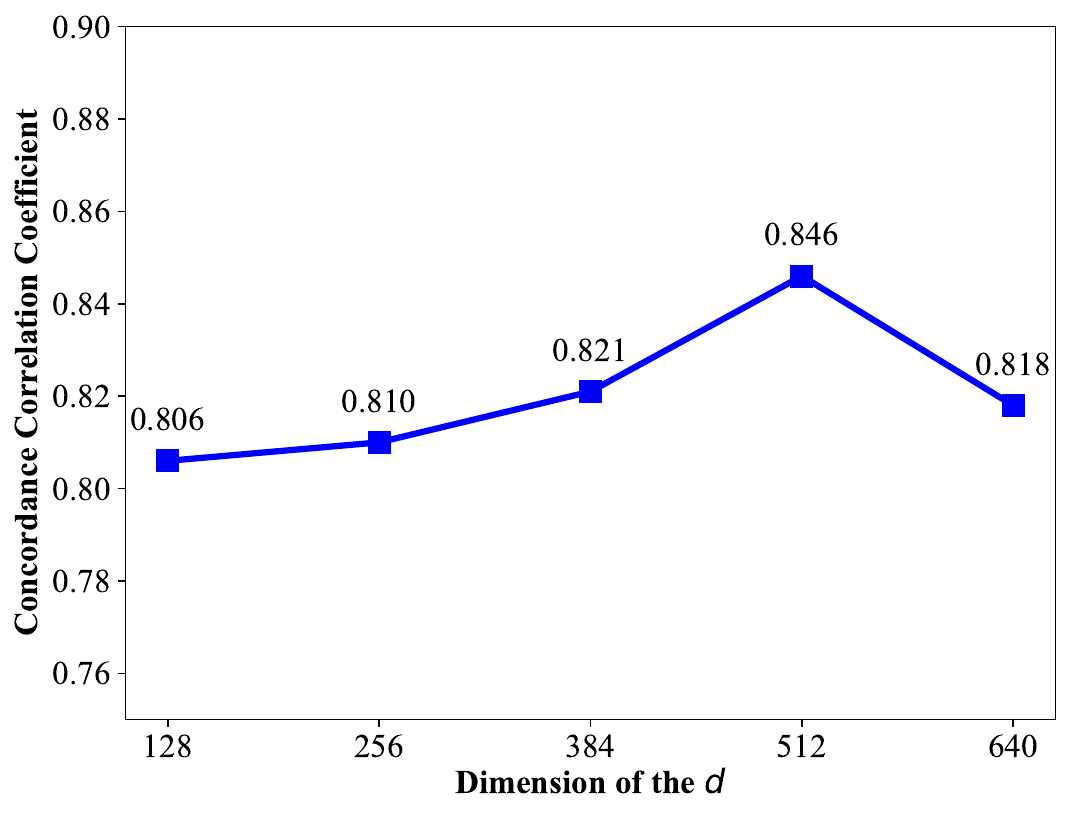}
    \caption{Ablation on the unified mapping space dimension $d$.}
    \label{fig:d_space}
\end{figure}

\begin{figure*}[!h]
  \centering
  \includegraphics[width=0.92\linewidth]{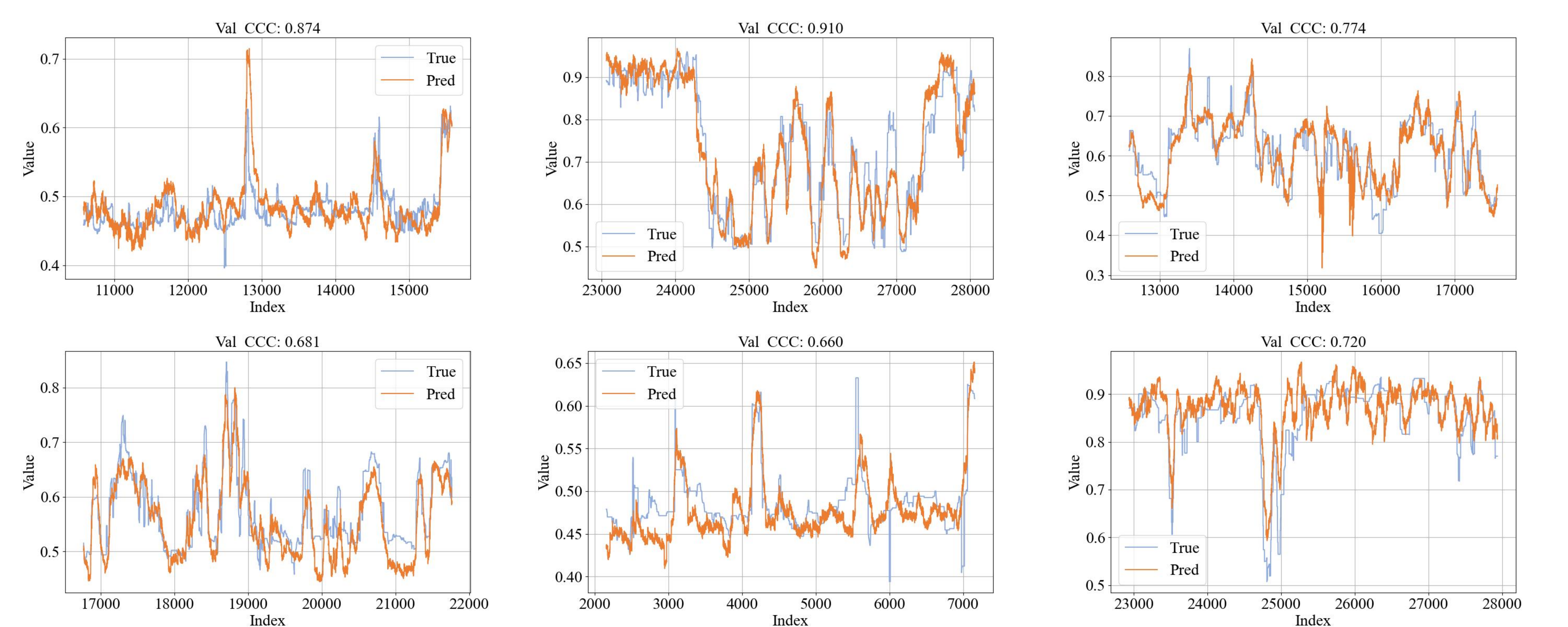}
  \caption{Using our method for real-time fitting, the selected interval length is fixed at 5000 samples.}
  \label{fig:label_sample}
\end{figure*}

\noindent \textbf{Ablation on Dialogue-Aware Encoder.}
Regarding the performance enhancement attributed to our proposed Dialogue-Aware Encoder (DAE), an ablation study was meticulously executed, with results summarized in Table \ref{tab:tab1}. The data unequivocally indicate that the integration of DAE yielded an improvement of 0.008 in the CCC on the Noxi Base dataset, thereby validating the contribution of DAE to the model's predictive accuracy. Moreover, when DAE was synergistically paired with MGF, the model's performance was further elevated, achieving an additional gain of 0.020 in the CCC. This incremental enhancement not only underscores the effectiveness of the DAE module but also highlights its complementary benefits when integrated with MGF, thus reinforcing the holistic efficacy of our proposed framework. 
Additionally, to verify that the benefits of the DAE module are not due to an increase in parameter count, we compared models with and without the DAE module while increasing the model depth to keep the parameter count consistent. As shown in Table \ref{your-label}, the model with the DAE module outperforms the model without it on the validation set, with a 0.025 increase in CCC, demonstrating the effectiveness of the DAE module.

\noindent \textbf{Ablation on the Project Dimension $d$.}
The size of the spatial dimension of the unified mapping module in the model framework directly influences the model's learning ability and performance. We have explored various sizes for the dimension 
$d$. As shown in Figure \ref{fig:d_space}, regarding the validation set CCC score for the NoXi Base data, the default size of 512 for $d$ in our model is found to be optimal. Dimensions that are too small result in inadequate expressive power for the model, whereas excessively large dimensions lead to overfitting. Therefore, dimensions that are either larger or smaller than 512 lead to a reduction in the model's performance.

\subsection{Presentation of Final Results}
Finally, our model framework, adjusted to the optimal settings, was applied to the test sets of all three datasets and compared with the official baseline models ~\cite{muller2024multimediate}. As shown in Table~\ref{tab:final_test}, our model achieved a state-of-the-art result on the NoXi Base test set with a 0.76 CCC. Additionally, it demonstrated superior performance on the NoXi-Add (0.67) and MPIIGI (0.49) datasets, showcasing the robustness of our method. Furthermore, our approach surpasses the official baseline by 0.23 on the global average CCC, indicating its effectiveness. To visually highlight our model's performance, we present comparisons of predicted and true labels on the NoXi Base validation set, as shown in Figure~\ref{fig:label_sample}. These visualizations show that our predictions largely capture the fluctuations in participants' engagement over time.


\begin{table}[h!]
  \caption{Final CCC scores on various test datasets. NoXi Base, NoXi-Add, and MPIIGI represent the NoXi Base, NoXi Additional Language, and MPIIGroupInteraction datasets, respectively. Global: global average CCC across all datasets.}
  \label{tab:final_test}
  \scalebox{0.88}{
  \begin{tabular}{ccccc}
    \toprule
    \makecell[c]{Team }&\makecell[c]{NoXi Base}& \makecell[c]{NoXi-Add}& \makecell[c]{MPIIGI}& \makecell[c]{Global}\\
    \midrule
    USTC-IAT-United & 0.72 & 0.73 & 0.59 & 0.68 \\
    AI-lab          & 0.69 & 0.72 & 0.54 & 0.65 \\
    \textbf{HFUT-LMC (Ours)} & \textbf{0.76} & \textbf{0.67} & \textbf{0.49} & \textbf{0.64} \\
    Syntax          & 0.72 & 0.69 & 0.5  & 0.64 \\
    ashk            & 0.72 & 0.69 & 0.42 & 0.61 \\
    YKK             & 0.68 & 0.66 & 0.4  & 0.58 \\
    Xpace           & 0.7  & 0.7  & 0.34 & 0.58 \\
    nox             & 0.68 & 0.7  & 0.31 & 0.56 \\
    SP-team         & 0.68 & 0.65 & 0.34 & 0.56 \\
    YLYJ            & 0.6  & 0.52 & 0.3  & 0.47 \\
    MM24 Baseline \cite{muller2024multimediate}   & 0.64 & 0.51 & 0.09 & 0.41 \\
    
  \bottomrule
\end{tabular}}
\end{table}

\section{Conclusion} 
\label{conclusion}

In this paper, we introduce a novel, language-independent Dialogue-Aware Transformer (DAT) framework that exclusively utilizes audio-visual inputs to estimate human engagement in conversations. The Modality-Group Fusion (MGF) employed in this study effectively fuses each participant's audio-visual features independently, significantly enhancing the model's performance and robustness. Furthermore, the inclusion of the Dialogue-Aware Encoder (DAE), which integrates behavioral and conversational information from both the target participant and their conversational partners, significantly augments the target participant's data, effectively mitigating bias in the model. With 
integrating MGF and DAE, our approach not only achieves a CCC of 0.76 on the NoXi Base test set but also surpasses the established baseline by an average of 0.23 CCC across NoXi Base, NoXi-Add and MPIIGI test sets. We believe that DAT will serve as a solid baseline and contribute significantly to engagement estimation. In the future, we will explore more effective methods to leverage partner information to further enhance engagement estimation in dialogues.

\begin{acks}
This work was supported by the National Natural Science Foundation of China under Grants 62202139 and 61932009, and in part by The University Synergy Innovation Program of Anhui Province under Grant GXXT-2022-038.
\end{acks}

\bibliographystyle{ACM-Reference-Format}
\balance
\bibliography{mybib}










\end{document}